\documentclass[usenatbib]{mn2e}

\usepackage[dvips]{graphicx}
\def\msun{M$_\odot$}

\title[]{Multiple Populations in Globular Clusters: The Possible Contributions
  of Stellar Collisions}

\author[Alison Sills and Evert Glebbeek] {Alison Sills and Evert
  Glebbeek\thanks{Email: asills@mcmaster.ca,
    glebbeek@mcmaster.ca}\\ Department of Physics and Astronomy,
  McMaster University, Hamilton, Ontario L8S 4M1, Canada}

\begin{document}
\maketitle

\begin{abstract}

Globular clusters were thought to be simple stellar populations, but
recent photometric and spectroscopic evidence suggests that the
clusters' early formation history was more complicated. In particular,
clusters show star-to-star abundance variations, and multiple
sequences in their colour-magnitude diagrams. These effects seem to be
restricted to globular clusters, and are not found in open clusters or
the field.  In this paper, we combine the two competing models for
these multiple populations and include a consideration of the effects
of stellar collisions. Collisions are one of the few phenomena which
occur solely in dense stellar environments like (proto-)globular
clusters. We find that runaway collisions between massive stars can
produce material which has abundances comparable to the observed
second generations, but that very little total mass is produced by
this channel. We then add the contributions of rapidly-rotating
massive stars (under the assumption that massive stars are spun up by
collisions and interactions), and the contribution of asymptotic giant
branch stars. We find that collisions can help produce the extreme
abundances which are seen in some clusters. However, the total amount
of material produced in these generations is still too small (by at
least a factor of 10) to match the observations. We conclude with a
discussion of the additional effects which probably need to be
considered to solve this particular problem.
\end{abstract}

\section{Introduction}

Globular clusters have long been viewed as the epitome of simple stellar
populations. Their stars have a common age, a common
distance, and a common metallicity; there is no interstellar gas and
little else to get in the way of studying the stars directly. These
systems are the closest we can come to a ``controlled experiment'' in
stellar astrophysics, and as such they have proved incredibly valuable
for studies of both stellar evolution and stellar dynamics in the past.

However, in recent years, cracks have been appearing in this simple
picture. Both photometric and spectroscopic studies of clusters have
started to unearth puzzles and inconsistencies. At first, these
problems were thought to be an oddity in one particular cluster, or
evidence of some details of stellar physics that we didn't quite
understand. Over the last five years or so, however, it is becoming
clear that our picture of a globular cluster needs to change. They
cannot have formed instantly out of a single molecular cloud, removing
all their leftover gas immediately, and then evolved passively for
the next 10 billion years. Their history is more complicated. 

\subsection {Observational Background}

Hints that something strange was going on in globular clusters came
first from spectroscopic studies of their red giants. For an excellent
review, see \citet{2004ARA&A..42..385G}. The general results have not
changed since that review was written, although we now have
observations of more stars per cluster, and more stars observed with
high-resolution spectra. Most globular clusters have a constant iron
and iron-peak element abundances, with the notable exception of
$\omega$ Centauri \citep{1975ApJ...201L..71F} and hints of a very
small spread in M22 and M92 \citep{2009A&A...505.1099M,
  1998AJ....115..685L}. However, it has been known since the late
1970s that lighter metals, particularly carbon, nitrogen and oxygen,
do vary from star to star in many clusters
\citep{1978ApJ...223..487C}.  Other light elements also show
star-to-star variations in clusters, including Na, Al and Mg.  The
most striking piece of evidence to date that this abundance variation
phenomenon occurs in all clusters is the spectroscopic study of sodium
and oxygen by \citet{2009A&A...505..117C,2009A&A...505..139C}. A
general anti-correlation is seen, with oxygen-depleted stars having
higher sodium abundances.  Aluminum and magnesium have been studied in
fewer clusters, but they also show similar trends
\citep{1996AJ....112.1517S,2009A&A...505..139C}, with a large range in
aluminum, a smaller spread in magnesium, and some evidence for
high-aluminum stars having lower magnesium abundances.

\citet{2008arXiv0811.3591C} also looked for correlations between the
extent of the Na-O anti-correlation and cluster properties. The
strongest correlation was between the extent of the correlation and
the maximum temperature of stars on the horizontal branch, in the
sense that clusters with very hot (or blue) horizontal branches had a
large spread in Na and O. They also found a weaker trend with the
total mass (or magnitude) of the cluster, and a trend with galactic
orbit, in the sense that clusters which have spent more of their
lifetime in the outskirts of the halo and not interacting with the
galaxy, have a larger spread in abundances.

One of the main reasons that globular clusters have been so useful to
stellar astrophysics is that their colour-magnitude diagrams are very
clean. Other than the blue straggler stars and a few other unusual
objects, the stars fall onto a single isochrone \citep[see
  e.g.][]{2007AJ....133.1658S}. Again, the exception has been $\omega$
Centauri. Its colour-magnitude diagram shows a large amount of
structure beyond a single age/composition isochrone. The spread of the
giant branch of this cluster was understood in the context of the
measured differences in iron abundance of the stars, and so it did not
come as much of a surprise to the community when the main sequence of
this cluster was found to contain more than one sequence as well
\citep{2004ApJ...605L.125B}. However, the most intriguing result from
this detailed study of $\omega$ Centauri was the determination that
the iron abundance of the bluest main sequence was not, as one would
expect, the lowest in the cluster, but was in fact the highest
\citep{2005ApJ...621..777P}. The only possible way to reconcile the
spectroscopic abundances with the photometric information was to infer
a high helium content for these stars, perhaps even as high as
Y=0.4. Because the abundance anomalies are found in light elements
only but not iron, researchers have been casting this as a problem of
'pollution' from an early generation of stars in the cluster.

The globular cluster community finally came to realize that multiple
populations were ubiquitous when \citet{2005ApJ...631..868D} discovered
an intrinsic spread in the main sequence NGC 2808, which was later
confirmed to be {\em three} separate main sequences
\citep{2007ApJ...661L..53P}. The turnoff region in this cluster is
quite tight without much evidence for a spread, but starting about one
magnitude below the turnoff, the sequences become obviously
separated. 

At about the same time, we saw the first CMDs showing multiple
subgiant branches (e.g.the ACS observation of NGC 1851
\citep{2008ApJ...673..241M}). Similar observations have since been
seen in M22, NGC 6388, and M54 \citep{2009arXiv0902.1422P} and many
intermediate age clusters in the LMC and SMC
\citep{2008ApJ...681L..17M,2009A&A...497..755M}. NGC 1851 is an
interesting case, is that it does {\em not} show any evidence for a
splitting of the main sequence, even in very careful observations
including proper-motion cleaning of the CMD. It does, however, show a
split in its giant branch when observed in $U-I$
\citep{2009ApJ...707L.190H} which is not seen in $V-I$, highlighting
the importance of observing in multiple bands.

Understanding the horizontal branch (HB) morphology in globular
clusters has been a problem for decades, commonly referred to as the
`second parameter problem.' Because the position of HB stars in the
CMD is sensitive to both helium and metal abundances, they are a very
useful population with which to discuss both pollution and multiple
populations \citep[e.g.][]{2008MNRAS.390..693D}.

\subsection{Possible Explanations}


Many of the models to date have attempted primarily to understand the
source of the pollution. The work of \citet{2007A&A...470..179P}
showed that the general abundance patterns could be explained by
invoking the products of hot hydrogen burning, at a temperature of
70-80 $\times 10^6$ K. At these temperatures, the hydrogen is burned
in a series of cycles -- the standard CNO cycle, as well as the
neon-sodium cycle and the aluminum-magnesium cycle. Conveniently, the
second two cycles work to give the abundance patterns that are needed:
higher sodium, lower Mg and higher Al. This processed material must be
removed from the star before helium burning can proceed. Otherwise,
the helium will be converted to carbon and oxygen. We will not have
the large amount of helium that is required at the surface, and the
constraint of (almost) constant C+N+O abundance will be violated.

A number of polluters have been proposed. The leading contender is a
population of intermediate-mass (3-10 \msun) asymptotic giant branch
(AGB) stars \citep[for a review, see][]{2008MNRAS.391..354R}. These
stars can reach the appropriate high temperatures in their hydrogen
burning shells, and the processed material is brought to the surface
by the outer convection zone as it reaches into the burning shell
during the hydrogen burning portion of the thermal pulse phase. AGB
stars have strong but low-velocity winds, which means that any mass
that is removed from the star has a good chance of remaining in the
cluster. Indeed, there is recent evidence that giant stars in this
mass range do show surface abundances which are consistent with the
necessary pollution \citep{2009A&A...504..845V}.

An alternative source of pollution is a population of fast-rotating
massive stars \citep{2007A&A...464.1029D}. Massive stars reach
sufficiently high temperatures in their hydrogen-burning cores. These
stars also have substantial winds, but under normal circumstances, the
core regions are not exposed until very late in the stars' lives, well
past the helium-burning phase. If a star is rotating rapidly, however,
then meridional circulation and other rotational instabilities will
mix material from the core to the surface, bringing these
hydrogen-burning products up to a region where the wind can take them
away from the star. Also, if the star is rotating rapidly enough that
it is close to its break-up velocity, material can escape from the
equator of the star in a slow outflowing disc or slow wind. This
material in particular has a slow enough velocity that it will stay in
the cluster, and there are even indications that low mass stars could
form in situ in the disc.

Both AGB and fast-rotating massive star models have some difficulties
in explaining the pollution of globular cluster gas and the formation
of the second generation of stars. In both cases, the amount of
material that is required to form the second generation is quite
large. Observations of clusters such as NGC 2808 suggest that the mass
of the second generation is approximately equal to the first (within a
factor of 2-3 or so), which puts a limit on the amount of ejecta that
the first generation must produce. Early versions of both models
\citep{2004ApJ...611..871D,2007A&A...464.1029D} suggested non-standard
IMFs, heavily weighted towards the polluter in question, but these
IMFs are difficult to justify. Current versions suggest that in fact
the first generation needed to be significantly larger than what is
observed today, and that the cluster needs to have lost approximately
90\% of its first generation, while retaining all of the gas and the
second generation of stars
\citep{2009A&A...499..835V,2008A&A...492..101D}. Both models also
require a certain amount of dilution of the polluted ejecta with
primordial gas, in order to match the observed abundances of light
elements, including lithium.

The work of \citet{2009ApJ...697..275P} looked at the effect of both
enhanced helium abundance, and enhanced C+N+O abundance, on isochrones
appropriate for globular clusters. They confirm that multiple main
sequences are caused by a change in helium abundance at constant metal
abundance (both iron and light elements). Multiple subgiant branches,
on the other hand, are caused by a difference in total C+N+O abundance
at constant helium, constant iron, and constant
age. \citet{2009ApJ...707L.190H} show that the split subgiant and
giant branches of NGC 1851 are best explained in both $U-I$ and $V-I$
by a small change in both helium and metal abundance. The alternative
explanation is a difference in age of the two populations of $\sim$1
Gyr and constant abundances. It is, however, difficult to explain a
delay in star formation of such a long time, and difficult to
understand where the gas for the second generation came from.

Only a few groups have tried to put together the entire scenario. The
most successful is the work of \citet{2008MNRAS.391..825D}, which
incorporated hydrodynamic simulations to study the flow of the AGB and
supernovae ejecta in the vicinity of the proto-cluster. They find that
the ejecta collects in a cooling flow and returns to the core of the
cluster. They also look at a scenario in which pristine gas, which was
pushed out of the cluster vicinity by the supernovae and massive
stars, returns to the cluster after a few million years and mixes with
the ejecta. Similar to \citet{2007A&A...470..179P,
  2007MNRAS.379.1431D, 2009A&A...499..835V}, they find that the
combination of pristine gas and ejecta is necessary. They also find
that much of the first generation in the cluster may be lost if the
cluster is tidally limited.  Finally, both \citet{2008MNRAS.391..825D}
and \citet{2008A&A...492..101D} use N-body dynamical models to study
the subsequent evolution of this two-generation cluster and the mixing
of the two populations. They concur with the general results of
\citet{2007ApJ...662..341D} that, at the current time, a
two-generation cluster will not be dramatically different from a
single-generation cluster in terms of the dynamics and distribution of
stars.


\section{A Toy Model Calculation}

In this paper, we investigate a modified version of the scenarios
presented in the introduction.  First, we will consider the case where
{\em both} fast-rotating massive stars and asymptotic giant branch
stars contribute to the pollution of material which forms the
subsequent generation(s). Second, we include the possible effects of
stellar collisions.

Recall that the multiple populations phenomenon is only seen in
globular clusters and not in the field or in open clusters. Dense
globular clusters are one of the few places in the universe where
stellar collisions are expected to be prevalent
\citep{1976ApL....17...87H}. There are three relevant outcomes of
stellar collisions: runaway collisions, modification of the rotational
properties of stars, and creation of intermediate mass stars.

Dynamical calculations of dense stellar systems have shown that there
can be dozens of direct stellar collisions between massive stars
within the first few million years of the cluster's life
\citep{2004Natur.428..724P}. These runaway collisions involved stars
with a total mass of up to 1000 \msun~ and 
occur in any cluster which undergoes a core collapse within the first
3 Myr or so \citep{2006MNRAS.368..121F}
The runaway collision 
dramatically modifies the upper end of the initial mass function. The
most massive stars in the cluster will not follow their normal
isolated evolution, but will merge to create the collision
product. Therefore, there will be fewer supernovae and fewer neutron
stars and black holes created. Secondly, the evolution of these very
massive stars must include substantial mass loss, but the mass loss
history of the collision product is very different from that of the
sum of the stars which go into the collision. As shown by
\citet{2009A&A...497..255G}, the total amount of mass ejected by the
collision product can be up to a factor of $\sim$1.5 larger than that
ejected by the collision parents individually.

At the same time that the runaway collision is happening, other
collisions will also be going on in the cluster. Some of them will
produce massive collision products, and others will be between very
small and very large stars, which will produce rapidly-rotating stars.
Unless two stars collide exactly head-on, most of angular momentum of
the original trajectory is deposited into the collision product,
producing an object which is spinning rapidly
\citep{1996ApJ...468..797L}. Many of them could be rotating with a
substantial fraction of their break-up velocity. Therefore, the
population of fast-rotating massive stars should be larger in a
cluster with a high collision rate.

And finally, collisions between low and intermediate mass stars could
increase the number of stars in the 3-10 \msun~ range. Some of the
original stars in this range will have been removed due to collisions
themselves, and so determining the total number of intermediate mass
stars is not a trivial calculation. However, these stars will not
segregate quickly enough to become part of the sub-cluster which
creates the runaway collision.  This population does not start
contributing to the gas of the subsequent generation until the stars
reach the AGB phase, around 30-100 Myr after the formation of the
cluster.

To try to get a handle on the abundances of the gas which could
possibly form the second and third generations of stars, we calculate
a very simple chemical evolution model based on the scenario described
above. We also follow the mass function of stars as they form, evolve
to remnants, and are modified by collisions. We begin with a gasless
star cluster at t=0. We assume some typical quantities for a globular
cluster: the total mass of stars in the system is $1 \times 10^5$
\msun, its half-mass radius $r_h$=3 pc, and the stellar velocity
dispersion is 10 km s$^{-1}$. According to the simulations of
\citet{2006MNRAS.368..121F}, a runaway collision will occur if the
cluster has an initial W$_0$ (the dimensionless central potential of a
King model) of 8 or higher, corresponding to a concentration parameter
$c = \log(r_t/r_c)$ = 2 or higher \citep[][figure
  4.10]{2008gady.book.....B}. For King models, $r_h/r_t$ is
approximately constant, with a value of 0.12 at $c=2$
\citep[][p. 16]{1987degc.book.....S}, and therefore this globular
cluster must have a core radius $r_c = 0.25$ pc, tidal radius $r_t$ =
25 pc, and a central density of 2.5 $\times 10^5$ \msun/pc$^3$ in
order for this scenario to be valid. These quantities are not
unreasonable for dense globular clusters seen today. We assume the
stars are formed with a Salpeter mass function between 0.1 and 120
\msun~ and have Z=0.001, Y=0.24 and an alpha-enhanced abundance
mixture with [$\alpha$/Fe]=+0.4.

Early in the cluster's life, we assume that one runaway collision
occurs. We remove each of the stars which participate in this
collision from the initial mass function. The yields from this
collision are based on the low-metallicity sequence in the
calculations of \citet{2009A&A...497..255G}.  The original calculation
used a simplified nuclear network that could only follow a small
number of species. We have recalculated the yields of the
\citet{2009A&A...497..255G} calculation using the nucleosynthesis
routines of \citet{2005MNRAS.360..375S} with reaction rates from
\citet{1999NuPhA.656....3A} and \citet{2006PhRvC..73b5802H}.  The
reaction network includes proton, neutron and $\alpha$ captures on
stable isotopes up to $^{34}$S and some iron group elements.  

Material leaves the runaway collision products in two ways. First,
during each collision, the product loses a few percent of its total
mass because of the energy of impact of the collision itself. The
amount of each element ejected in this way, summed over all collision
in the runaway sequence, are listed in table \ref{runaway} as
`ejecta'. Between collisions, the collision product is evolving as a
relatively normal massive star, and exhibits strong mass loss. In our
calculations, we assume that this loss is from a radiative wind,
although LBV-like mass loss may also be important.  The total amount
of material lost by the runaway collision product in winds is listed
in table \ref{runaway}. There are two wind channels, listed
separately: the wind of the collision product in between collisions {`Wind'),
and the wind of the collision product during its remaining lifetime
\emph{after} the last collision (`Rem. wind').  These winds should
leave the parent star with velocities which are typical of O stars and
luminous blue variables, which range from a few hundred to a few
thousand km s$^{-1}$. In a dense stellar system with many other
massive stars present, it is plausible that these winds will collide
with the winds of other stars, shock, and slow down sufficiently to
remain trapped in the centre of the cluster and contribute to the
second stellar generation. We assume that is the case initially, but
also perform the calculation under the assumption that all the wind
material is lost from the cluster.

As in \citet{2009A&A...497..255G} we were only able to follow
the evolution of the collision product about half way through core helium
burning due to numerical difficulties during this evolution phase.
We calculate the final remnant mass of the collision product as described in
\citet{2009A&A...497..255G} and estimate the yields for the remaining
core helium burning lifetime by assuming that the remainder of the envelope
is ejected without undergoing additional processing.
The results are listed in table \ref{runaway} under `Rem. remaining'.

\begin{table}
\centering
\caption{Abundances, in mass fraction, from the runaway collision. The total amount of mass in each component is given. \label{runaway}}
\begin{tabular}{lccccc}
\hline
Element & Ejecta &   Wind & Rem. wind & Rem. remaining\\
       &  145.1 \msun &  225.4 \msun &  84.1 \msun & 84.2 \msun \\
\hline
$^{1}$H     & 5.97e-01 & 3.09e-01 & 1.21e-01 & 5.70e-02 \\
$^{4}$He    & 4.02e-01 & 6.90e-01 & 8.78e-01 & 9.50e-01 \\
$^{12}$C    & 1.89e-05 & 1.02e-05 & 8.78e-06 & 7.48e-06 \\
$^{13}$C    & 2.16e-06 & 2.24e-06 & 2.31e-06 & 2.12e-06 \\
$^{14}$N    & 6.66e-04 & 7.10e-04 & 7.09e-04 & 7.21e-04 \\
$^{16}$O    & 5.03e-05 & 1.79e-05 & 1.21e-05 & 7.40e-06 \\
$^{19}$F    & 2.09e-09 & 4.44e-10 & 1.56e-10 & 2.12e-11 \\
$^{20}$Ne   & 7.03e-05 & 4.31e-05 & 2.85e-05 & 2.41e-05 \\
$^{23}$Na   & 2.20e-05 & 1.29e-05 & 3.94e-06 & 2.16e-06 \\
$^{24}$Mg   & 3.60e-05 & 7.50e-05 & 9.75e-05 & 9.97e-05 \\
$^{25}$Mg   & 3.82e-07 & 7.65e-08 & 3.14e-08 & 3.31e-09 \\
$^{26}$Mg   & 2.31e-06 & 7.35e-07 & 2.37e-07 & 1.36e-07 \\
$^{26}$Al   & 1.95e-06 & 1.20e-06 & 7.16e-07 & 5.73e-07 \\
$^{27}$Al   & 4.01e-06 & 5.62e-06 & 6.11e-06 & 6.24e-06 \\
\hline
\end{tabular}
\end{table}

This particular runaway is only one possible combination of parent
stars, impact velocities, etc. It was chosen in the study of
\citet{2009A&A...497..255G} to be representative of the collisions
seen in N-body simulations. The details of exactly how much mass, and
the exact composition of the ejecta, will change slightly if the
details of the runaway collision changes. This must be kept in mind
when interpreting the results of the chemical evolution calculations
presented below. A comparison of the three runaway collision sequences
studied in \citet{2009A&A...497..255G} show that the yields are
consistent to within 10-20\% depending on the element. The total
amount of mass which can be released varies by up to a factor of two
(up to approximately 1000 \msun). As we will discuss below, the
results do not change dramatically if we adopt the upper limit of
these values.

While the runaway collision is going on, other stellar collisions will
spin up some of the massive stars in the cluster. Following
\citet{2004ApJ...604..632G}, we assume that the massive stars in the
cluster segregate towards the centre, and form a decoupled dynamical
cluster which undergoes core collapse. This cluster, until very close
to the time of core collapse, has the same core radius and central
density as the original cluster. However, the mass function becomes
much more weighted towards massive stars, so that the average stellar
mass in this sub-cluster is more like 20 \msun~ (compared to $\sim$
0.35 \msun~ for a normal cluster). Using these parameters and the
equation for the average time between collisions from
\citet{1989AJ.....98..217L}, we find that approximately 1100 stellar
collisions should occur during this first 5 Myr in this initial
sub-cluster of stars. There are approximately 200 stars in this
cluster with masses above 20 \msun, forming approximately 35\% of the
stars in the sub-cluster. Therefore, it is likely that every high-mass
star has undergone at least one stellar collision.

Following \citet{2007A&A...464.1029D}, we define ``rapidly rotating''
to mean having a rotation rate that is at least 80\% of the critical
rotation rate. We expect that some fraction of massive stars are
primordial fast rotators (i.e. they were born that way and not spun
up by collisions). The models of \citet{2007A&A...464.1029D} assume
that all massive stars are rotating rapidly enough to contribute
polluted material to the second generation. Observations of stars in
young clusters \citep[e.g.][]{2006A&A...457..265D} suggest that
cluster stars are more rapidly rotating than those in the field, and
that the fraction of rapid rotators is more like 20\%. Initially, we
make the extreme assumption that every collision will turn a slow
rotator into a rapid rotator. This assumption is unlikely to be
completely correct, as the amount of angular momentum which can be
added to a high-mass star depends on the mass of the impactor, its
velocity and its position of impact on the high mass star. However,
the highly collisional environment of the mass-segregated sub-cluster
means that high mass stars will likely undergo more than one
collision. The net effect of these collisions will be to spin up the
population.

We use the yields of \citet{2007A&A...464.1029D} to determine the
contribution of this population to the ejected gas. We use the yields
of their case C reaction rates for 40, 60 and 120 \msun~ fast-rotating
stars. These yields come from a set of reactions in which the
reactions involving $^{20}$Ne through $^{27}$Al are set to their
experimental upper or lower limits in such a way to produce the most
favourable set of yields for this work. However, we do not use the
yields from the models in which the $^{24}$Mg(p,$\gamma$) reaction
rate is increased by a factor of 1000 at $5\times10^7$ K (case
D). This fudge was required to produce the lower range of the observed
[Mg/Al] values in NGC 6752. However, for this toy model, we prefer to
use the 'standard' set given by \citet{2007A&A...464.1029D}.  The
models presented in that paper assumed a metallicity of Z=0.0005 and
are initially alpha-enhanced. The yields for the runaway collisions
and the AGB stars are for Z=0.001, so we should be using different
yields to be consistent. Unfortunately higher metallicity yields for
fast-rotating massive stars are not available. We do not expect our
results to be dramatically different if we had used other reasonable
yields or models.
 
Like the runaway collisions, these fast-rotating massive stars also
lose material in two ways. One, a slow wind is present, primarily from
the equator of the stars, as the surface material reaches an angular
velocity higher than the local stellar escape velocity. This material
will remain in the cluster, and those yields are assumed to be those
up to the time the star has reached the end of central H burning,
following \citet{2007A&A...464.1029D}. These stars also have a fast
wind, with velocities typical for O and B stars (a few hundred km
s$^{-1}$), and is dominant between the end of central H burning and
the time when the helium-burning products appear at the surface of the
fast-rotating stars.  We make the same assumptions as for the winds
of the runaway collision: initially we assume these winds shock and
remain in the cluster, and then we will present a calculation in which
they are assumed to be removed entirely from the cluster.

The time of the first supernova is set by the most massive star which
is not involved in a runaway collision, or the lifetime of the runaway
itself, whichever comes first. At this time, which is approximately 5
Myr, we assume that all the gas in the cluster forms the second
generation of stars, and any subsequent gas is removed from the
cluster until the AGB stars begin to contribute. For this reason, we
neglect the contribution of rapidly rotating 20 \msun~ stars, as their
main sequence lifetimes are closer to 10 Myr.  We calculate the total
mass in ejecta from both the runaway collision and the fast
rotating massive stars, under the assumption that both fast and slow
winds are retained in the cluster. We find that we have 3459 \msun~ of
material, which we assume forms the second generation of stars with
100\% efficiency. We populate a Salpeter IMF from 0.1 to 120 \msun, as
was done with the initial cluster.

The abundances of this second generation are quite extreme. In Figure
\ref{fig:Yhist}, we show the helium abundance of each generation,
labelled as ``primordial'', ``runaway \& FRMS'' for this second
generation, and ``all AGB'' and ``high mass AGB'' for different
possibilities of the third generation, to be discussed below. This
figure also shows the calculated number of low mass stars, with masses
less than 0.8 \msun, which are expected to be observable members of
the cluster today. Note that we plot the logarithm of the number of
stars.  The runaway collision, which contributes only 539 \msun~ of
material to the cluster, is strongly enhanced in helium, with an
overall Y of 0.68. The fast-rotating massive stars are also strongly
helium enhanced, with Y=0.43 for the 2920 \msun~ ejected. This results
in an overall helium abundance for this first generation of
Y=0.47. This is much higher than is inferred in even the most extreme
globular cluster second generation.

The abundances of other elements, however, match the observations
reasonably well. Figures \ref{fig:NaOdiagram} and
\ref{fig:AlMgdiagram} compare the calculated abundances of each
generation or pollution mechanism, as labelled, to the
observations. The shaded region on each diagram encompasses most of
the observed abundance trends, as taken from the figures in
\citet{2009A&A...505..117C} and \citet{2009A&A...505..139C}.  The
runaway collisions alone seem to be the only way to reach significant
oxygen depletion ([O/Fe] $\sim$ -0.9) and sodium enhancement ([Na/Fe]
$\sim$ 1.0). The fast-rotating massive stars do deplete oxygen and
enhance sodium as well, although to a lesser extent. Because the
runaway collision produces so little mass, the overall abundances of
the second generation are oxygen-poor and sodium-rich at levels which
are consistent with all but the most extreme observations in clusters
(e.g. in NGC 2808).

Magnesium and aluminum abundances from this second generation are also
consistent with most of the observations. Again, none of our models
are able to reproduce the most extreme populations (the high aluminum,
low magnesium portion of the diagram). It should be noted, however,
that there are very few stars in that region and they are all in NGC
2808; the bulk of the stars in other clusters lie in the vertical
region of this plane, and we argue that magnesium depletion needs to
be confirmed in more clusters before this diagnostic is used to rule
out particular models. We also note that this generation does not
produce very much aluminum, despite the interestingly high helium
yields and the good agreement with the bulk of the Na-O observations.

Next, we consider the likelihood that collisions in this early
stage will produce additional intermediate-mass AGB stars later in the
cluster's life. We used the properties of the entire cluster core
rather than that of the sub-cluster of massive stars because the low
to intermediate mass stars that we were considering here have not had
time to mass segregate. Using equation 13 of
\citet{1989AJ.....98..217L}, we found that only 4 collisions should
have occurred in the first 5 Myr. Even if we assume that the cluster
conditions remain the same until the stars of this mass reach the AGB
phase (about 30 Myr), only a few dozen collisions will occur, and only
approximately 5\% of those will produce stars with masses between 3
and 8 \msun. Therefore, we will neglect the contribution of
collisionally-created AGB stars. 

Now we calculate the contribution of the AGB stars from both the first
and second stellar generations. We use yields for Z=0.001 for both the
first and second generation populations \citep{2008A&A...479..805V,
  2009A&A...499..835V}. These yields will be incorrect for the AGB
stars formed from the ejecta of the runaway collision and fast
rotating massive stars, as the helium and light element content of
these stars is initially increased. The structure of helium-rich stars
is different than that of normal stars, and the hydrogen-burning
cycles are dependent on the abundances of the catalyst
elements. Therefore, we should not simply scale the yields from the
normal AGB stars to estimate the yields of high-helium stars. The
helium abundances will be higher than given by the normal star yields
because even unprocessed material will be enriched in helium. However,
the number of second-generation AGB stars is low ($\sim$ 65), and so
we will take the conservative assumption that their yields are the
same as the first generation population. In addition, helium-rich AGB
stars have lifetimes which are $\approx$ 50-70\% shorter than stars of
the same mass but normal helium. In our simple model, this does not
affect our results because we simply sum up all the contributions from
AGB stars with lifetimes shorter than our cutoff. However, a more
detailed model will need to include this lifetime effect.

First, we assume that all AGB stars between 3 and 6 \msun~ contribute
to the material which forms the third generation.  The AGB stars
produce almost twice as much material as the massive stars (6084
\msun). The helium abundance of this population is Y=0.29, which is
higher than the standard value but still not as high as the Y=0.4
inferred for clusters such as NGC 2808. The AGB ejecta is slightly
enhanced in oxygen compared to the initial value ([O/Fe] = 0.55, up
from 0.4), and is significantly enhanced in sodium ([Na/Fe] = 1.04, up
from -0.2). The sodium enhancement is much larger than that seen in
most globular clusters, and the AGB stars alone do not produce stars
with low oxygen/high sodium values that are seen in globular
clusters. Dilution with primordial material reduce both the sodium and
oxygen abundances, but an oxygen depletion of approximately 1 dex is
impossible to accomplish with these AGB yields. Similarly, this
population produces some aluminum without much change in
magnesium. This population is labelled `all AGB' in figures
\ref{fig:Yhist} - \ref{fig:AlMgdiagram}.

However, 3 \msun AGB stars have lifetimes of over 300 Myr. This time
is long enough that SNIa may have started to pollute the cluster and
disrupt the gas. Also, the most massive AGB stars started losing their
mass after only $\sim$ 50 Myr, and it is not clear that this material
would have remained in the cluster, waiting for the ejecta of the
lower mass stars. It is more likely that the longest possible time for
the AGB ejecta to collect is more like 100 Myr, which is the lifetime
of a 5 \msun AGB star. If we restrict ourselves to only the most
massive AGB stars (5-6 \msun), then the sodium and oxygen yields are
more consistent with the observations, and in fact are very similar to
those from the runaway + FRMS population. Under this assumption, AGB
stars only contribute $\sim$ 2100 \msun~ to the new generation, an
amount of mass which is comparable to that of the first
generation. This population also produces more aluminum and shows a
very slight magnesium dilution. This population is labelled `high mass
AGB' in figures \ref{fig:Yhist} - \ref{fig:AlMgdiagram}.

If we combine the second and third generation, we have 9.5 $\times 10
^3$ \msun~ of material, or less than 10\% of the initial cluster. If
we compare the number of stars that will still be in the globular
cluster at current time (less massive than 0.8 \msun), the two new
generations have created 25 000 stars and there are approximately 2.6
$\times 10^5$ stars from the first generation. We are still required
to lose 90\% of the initial low-mass stars in order to have our
younger generations form half the cluster at the current day. These
numbers assume that all AGB stars contribute to the third generation,
which is the most generous assumption one can make about total mass,
but is almost certainly an overestimate as discussed above.

The other way to mitigate this mass problem is to allow primordial
material to mix into the gas which will form either the second or
third generation (or possibly both). While this will certainly help
boost the mass of that generation, it will also change the
abundances. In figures \ref{fig:NaOdiagram}
and \ref{fig:AlMgdiagram}, we have drawn lines of dilution for the
second and the two possible third generations. The amount of dilution
ranges from almost none near the points labelled by the polluters, to
a huge amount of mass as the line nears the primordial abundance. We
feel that it is more likely that the second generation would be
polluted than the third, and so we calculated the total amount of mass
needed to bring the helium abundance down to some values of
interest. For example, only 1500 \msun is needed to bring the helium
abundance to Y=0.4, but 77 000 \msun brings Y to 0.25. This is
comparable to the mass of the initial generation.

\begin{figure}
\begin{center}
\includegraphics[width=3.2in]{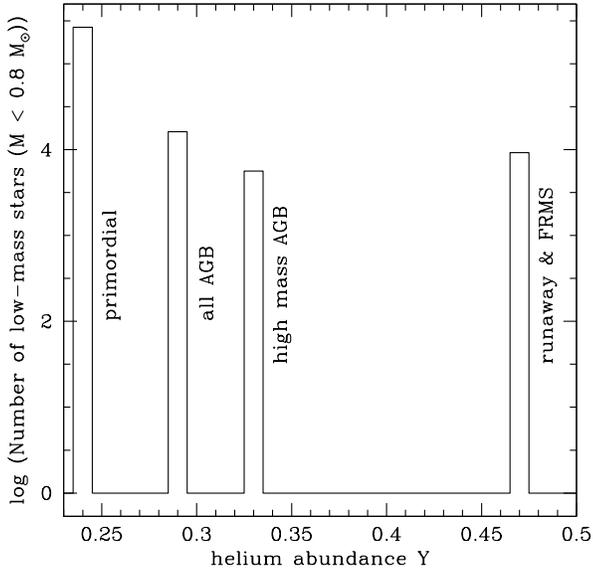}
\end{center}
\caption[]{Histogram of helium abundance of the three stellar
  generations, and the number of low-mass (i.e. observable) stars in
  each generation. The polluter(s) of each generation is marked:
  primordial abundance, all asymptotic giant branch stars, high mass
  asymptotic giant branch stars only, and runaway collision product \&
  fast-rotating massive stars. Note that the y axis is logarithmic.}
\label{fig:Yhist}
\end{figure}

\begin{figure}
\begin{center}
\includegraphics[width=3.2in]{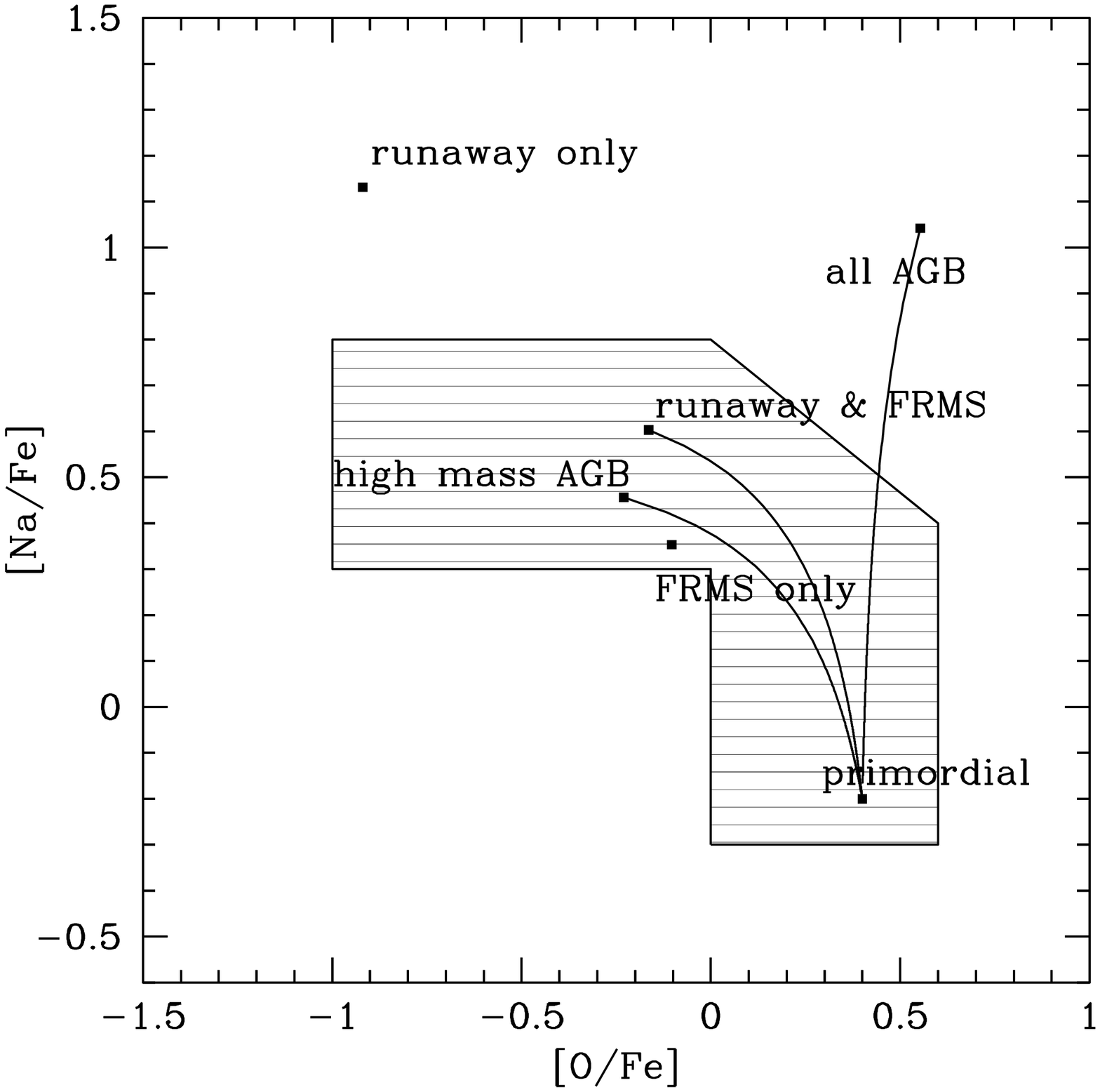}
\end{center}
\caption[]{Sodium vs. Oxygen abundances for the three stellar
  generations. The shaded region shows the extent of the observations,
  as taken from \citet{2009A&A...505..117C, 2009A&A...505..139C}. The
  abundances of the runaway collision products, the population of fast
  rotating massive stars, and the population of high-mass AGB stars
  only, are also marked. The curved lines connecting the generations
  to the primordial abundances show the effects of dilution.}
\label{fig:NaOdiagram}
\end{figure}

\begin{figure}
\begin{center} 
\includegraphics[width=3.2in]{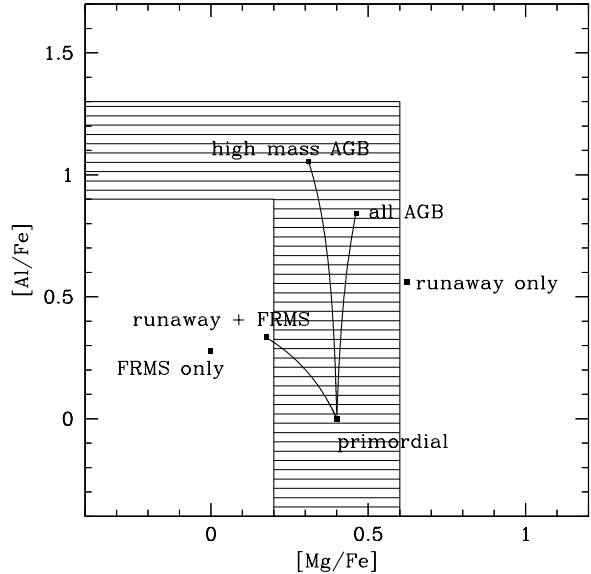}
\end{center}
\caption[]{Aluminum vs. Magnesium abundances for the three stellar
  generations. The shaded region shows the extent of the observations,
  as taken from \citet{2009A&A...505..117C, 2009A&A...505..139C}. The
  abundances of the runaway collision products, the population of fast
  rotating massive stars, and the population of high-mass AGB stars
  only, are also marked. The curved lines connecting the generations
  to the primordial abundances show the effects of dilution.}

\label{fig:AlMgdiagram}
\end{figure}

The previous results assume that both the ``slow'' and ``fast'' winds
from runaway collision products and fast-rotating massive stars
contribute to the second generation in the globular cluster. Fast
winds, with velocities between a few hundred and a thousand km/s, may
not be retained in the potential well of the initial cluster. We have
calculated the total mass of gas, and its abundances, using only the
slow winds from the two contributors to the second generation. As
expected, the total amount of mass available is reduced ($\approx$
1900 \msun~ instead of $\approx$ 3500 \msun), and the abundances are
different. In particular, the helium content is much lower (Y=0.321)
because the fast wind is thought to be occurring when the stars are
predominantly helium stars. The oxygen and sodium abundances are both
approximately 0.1 dex lower, while the aluminum and magnesium
abundances are lower by more like 0.15 dex. These abundances are still
consistent with the observed abundances in clusters. The oxygen and
sodium abundances are closer to the extreme values observed, while the
aluminum/magnesium values are closer to the primordial values.

For comparison, we have also calculated the expected contributions
from our three pollution processes if we neglect the effects of
stellar collisions entirely. In this case, runaway collisions do not
contribute at all. We assume that only 20\% of massive stars are
rotating rapidly enough to produce any polluting material (but we do
include both slow and fast winds from fast-rotating massive stars). We
assume that the other 80\% of massive stars are not rotating at all,
and we neglect any contribution of their winds to the second
generation. This is reasonable because their surface abundances have
not been modified by nuclear burning. The effects of AGB stars are
reduced as well, since we have much less mass with which to make
second-generation AGB stars. Therefore, the total amount of mass in
the subsequent generations is only 6500 \msun, a mere 6.5\% of the
initial mass of the cluster. Of this material, the bulk is from the
AGB stars of all masses, and so is very high in oxygen, sodium,
and aluminum, and only mildly enhanced in helium.

\section{Discussion and Conclusions}

In this paper, we take another theoretical look at the problem of
multiple populations in globular clusters. In particular, we tried to
address the observational constraint that this is specifically a {\em
  dense old cluster} phenomenon. We concentrated on the effects of
stellar collisions, which is one of the only physical mechanisms which
affects clusters much more strongly than anywhere else in the
universe. We also combined two previous models for multiple
populations -- the fast-rotating massive stars and the asymptotic
giant branch stars -- to determine if having more polluters would help
with either abundances or with the total mass in the subsequent
generations.

We included the effects of collisions in three ways: the impact of a
runaway collision, an increase in the number of fast-rotating
massive stars, and an increase in the number of intermediate mass AGB
stars. We also allowed for two subsequent generations. One is formed
within a few million years of the first cluster formation, from the
ejecta of runaway collisions and fast-rotating massive stars. The
other is formed approximately 100 million years later, from the ejecta
of the AGB stars. 


Our results are not substantially different from those of other groups
who are modeling the early evolution of multiple population
clusters. We still find that the mass produced by the polluters alone
is insufficient. Loss of a large fraction of the initial cluster,
dilution by primordial gas, or both, is required. However, stellar
collisions do produce material with abundances which are interestingly
extreme in all elements, and may well be a substantial contribution to
the extreme populations in the most massive clusters.

It is clear, from this work and others, that we do not yet have a
complete understanding of the multiple populations issue. Here, we try
to highlight a number of the largest remaining problems, starting with
those specific to our model and then discussing the puzzle more
generally.

So far, there have been no simulations of star formation in the
environment(s) envisioned here. Star formation typically occurs in
cores in molecular clouds, an environment which is cold, dark, dense,
and shielded from the outside universe. 
In the centre of a dense cluster, a
cloud of ejecta will be subject to the radiation field of the
first generation stars. 
No simulations of star
formation in a region with this kind of external radiation bath have
been done, and so we do not really understand the properties of the
second generation. We have assumed a Salpeter mass function for our
subsequent generations, for example, with a maximum mass of 120
\msun. It may be reasonable to postulate that high mass stars could
never form in the second generation (as was suggested by
\citet{2008MNRAS.391..825D}), which would mean that the second
generations have more low-mass stars then in our toy model. For a
Salpeter mass function between 0.1 and 120 \msun, 44\% of the mass is
found in stars more massive than 0.8 \msun. Under the assumption that
only stars of mass 0.8 \msun~ or less form in the subsequent
generations, we can increase the predicted mass of these generations
by approximately a factor of two.

If the cluster needs to be substantially more massive initially than
assumed in this paper, then we should consider how the different
polluter channels scale with cluster mass. For fast-rotating stars
and AGB stars it is reasonable to assume that the yields scale with
the total cluster mass. For the collision runaway this is not so
clear. The mass of the collisional runaway as a function of the
cluster mass has not been considered in detail in the literature. If
the mass of the collision runaway scales with cluster mass, then it
behaves as the other two polluter models. However, if the mass of the
collision runaway depends more steeply on the mass of the cluster,
then it will become relatively more important with increasing initial
cluster mass. In that case the number of stars that need to be lost
from the cluster can be smaller than when either of the other two
scenarios alone are considered. If the total mass of the runaway is
only a weak function of cluster mass the other polluter scenarios will
become relatively more important for increasing cluster mass.

In these models, we have neglected the effects of binary stars. We
know that globular clusters do have binaries, although work suggests
that the fraction may be lower than in the field and open clusters
\citep{2008AJ....135.2155D, 1997ApJ...474..701R}. Binary stars can
modify our toy model in a number of ways. Collisions between binary
stars are more likely than between single stars
\citep{1989AJ.....98..217L} because of the larger cross section of the
binary, and many of those interactions can result in more than two
stars merging. Binaries can also increase the likelihood of having
more than one runaway collision \citep{2006ApJ...640L..39G}.  There
are indications that massive stars have a higher binary fraction than
low mass stars in the field \citep{2006ApJ...640L..63L} and so it may
not be unreasonable to expect that interacting binary stars may have a
significant impact on the mass lost from massive stars. Following
suggestions by \citet{2009A&A...507L...1D},
\citet{2009arXiv0909.3431V} goes as far to suggest that interacting
massive binaries are responsible for all the pollution in globular
clusters, not fast-rotating massive stars or AGB stars. 

All models of multiple populations need to be refined substantially in
order to explain the cluster-to-cluster variations. While it is clear
that something is going on in almost every well-studied cluster, it is
not clear that we understand how that effect depends on the cluster
properties. In our scenario, we would argue that the
cluster-to-cluster variations were caused by different collision rates
early in the clusters' lives, perhaps driven by slightly different
formation conditions and initial densities. In fact, it may be that we
can use the extreme abundances produced in runaway collisions to
determine which clusters hosted a runaway all those years ago. Only
some clusters show evidence for extremely high Y values (Y $\approx$
0.4) and very extreme sodium-oxygen anti-correlations, such as $\omega$
Centauri and NGC 2808. Perhaps only those clusters were formed with
sufficiently high initial density to induce a runaway. A more detailed
study of the abundance patterns in the individual pollution mechanisms
may help disentangle these processes.

It has become clear that the epochs of globular cluster formation and
very early evolution are crucial pieces of the puzzle when
disentangling the present-day properties of these ancient objects. It
is also clear that our understanding of the dominant processes and
effects during these epochs is not as strong as we would
like. Learning about the early lifetimes of dense clusters is not easy
-- the stellar archaeology required is quite intricate. We are looking
at small changes in current surface abundances and brightnesses of
very old, low mass stars, and inferring a significant amount of action
involving more massive stars over 10 billion years ago. However, we
have learned a great deal about this phenomenon since it first was
identified only a few years ago, and progress will continue to be
made.

\section{Acknowledgments}
A.S. is supported by NSERC. A.S. and E.G. wish to thank the Kavli
Institute for Theoretical Physics and the Formation and Evolution of
Globular Clusters program for their hospitality, where this work was
first formulated. KITP is supported in part by the National Science
Foundation under Grant No. PHY05-51164.

\end{document}